\documentclass[11pt,aps,nofootinbib,showpacs]{revtex4}

\usepackage{epsf}
\usepackage{graphicx}
\usepackage{dcolumn}
\usepackage{bm}
\usepackage{amsmath}
\usepackage{amssymb}

\begin{document}
%



\renewcommand\({\left(}
\renewcommand\){\right)}
\renewcommand\[{\left[}
\renewcommand\]{\right]}
\newcommand{\pa}{\partial}
\newcommand{\dd}{{\rm d}}
\newcommand{\e}{{\rm e}}
\newcommand{\be}{\begin{equation}}
\newcommand{\ee}{\end{equation}}
\newcommand{\bea}{\begin{eqnarray}}
\newcommand{\eea}{\end{eqnarray}}
\newcommand{\half}{\frac{1}{2}}
               
\newcommand\mm{\,\mbox{mm}}
\newcommand\cm{\,\mbox{cm}}
\newcommand\km{\,\mbox{km}}
\newcommand\kg{\,\mbox{kg}}
\newcommand\TeV{\,\mbox{TeV}}
\newcommand\GeV{\,\mbox{GeV}}
\newcommand\MeV{\,\mbox{MeV}}
\newcommand\keV{\,\mbox{keV}}
\newcommand\eV{\,\mbox{eV}}
\newcommand\Mpc{\,\mbox{Mpc}}
\newcommand\yr{\,\mbox{yr}}
\newcommand\s{\,\mbox{s}}
\newcommand\Kelvin{\,\mbox{K}}
\newcommand\mpl{M_{\rm pl}}
\newcommand\Mpl{M_{\rm pl}}
\newcommand\msun{M_\odot}
\newcommand\mgut{M_{\rm GUT}}
\newcommand\bfx{{\bf x}}

\newcommand{\sn}{\tilde{N}}


\title{Resonant decay of flat directions}

\author{Marieke Postma~$^{a}$ and Anupam Mazumdar~$^{b}$} 
\affiliation{$^{a}$ The Abdus Salam International Center for 
Theoretical Physics, Strada Costiera 11, 34100 Trieste, Italy.\\
$^{b}$ CHEP, McGill University, Montr\'eal, QC, H3A 2T8, Canada. }
\date{July 22, 2003}

\begin{abstract}
We study preheating, i.e., non-perturbative resonant decay, of flat
direction fields, concentrating on MSSM flat directions and the right
handed sneutrino. The difference between inflaton preheating and
flaton preheating, is that the potential is more constraint in the
latter case. The effects of a complex driving field, quartic couplings
in the potential, and the presence of a thermal bath are important and
cannot be neglected.

Preheating of MSSM flat directions is typically delayed due to
out-of-phase oscillations of the real and imaginary components and may
be preceded by perturbative decay or $Q$-ball formation.  Particle
production due to the violation of adiabaticity is expected to be
inefficient due to back reaction effects.  For a small initial
sneutrino VEV, $\langle N \rangle \lesssim m_N/h$ with $m_N$ the mass
of the right handed sneutrino and $h$ a yakawa coupling, there are
tachyonic instabilities.  The $D$-term quartic couplings do not
generate an effective mass for the tachyonic modes, making it an
efficient decay channel.  It is unclear how thermal scattering affects
the resonance.

\end{abstract}


\maketitle

\renewcommand{\thefootnote}{\arabic{footnote}}
\setcounter{footnote}{0}



\section{Introduction}

During inflation the scale of quantum fluctuations is set by the
Hubble constant. Scalar fields which are light with respect to the
Hubble constant fluctuate freely, resulting in condensate formation.
Such condensates can play an important role in the evolution of the
universe. A condensate can affect the thermal history of the universe,
for example if its decay is accompanied by a large entropy
production~\cite{kt}. The baryon asymmetry in the universe may
originate from a condensate: Affleck-Dine baryogenesis utilizes a
scalar condensate along a flat direction of the supersymmetric
standard model~\cite{AD}, whereas Refs.~\cite{sneutrino} discuss
non-thermal leptogenesis from a decaying right-handed sneutrino
condensate.  Condensates can fragment into non-topological solitons,
called $Q$-balls, which may have implications for dark
matter~\cite{qballs,kasuya}.  In the curvaton scenario, quantum
fluctuations of a condensate generate the density perturbations
responsible for the observed CMB temperature anisotropy
\cite{early,sloth,curvaton}.

Field directions along which the effective mass vanishes, so-called
flat directions, are a generic feature of supersymmetric theories.
There are many directions in field space along which the potential
vanishes identically at tree level. The classical degeneracy along the
flat directions is protected from perturbative quantum corrections in
the supersymmetric limit by the non-renormalization theorem
\cite{grisaru}. The flatness is lifted by soft terms from
supersymmetry breaking. If the soft terms are sufficiently small
during inflation the fields along the flat directions will be
excited. Examples of flat directions in the MSSM are the $H_u L$ and
$\bar{u}\bar{d}\bar{d}$ directions~\cite{AD,gherghetta,flat}. Other
examples of flat directions are the right-handed sneutrino and moduli
fields.  The masses of pseudo-Goldstone bosons are protected by an
approximate global symmetry, and therefore can be kept naturally light
during inflation. The Peccei-Quinn axion is an example of a
pseudo-Goldstone boson that might condense during inflation.  We will
generically refer to fields parameterizing flat directions as {\it
flaton} fields, or simply {\it flatons}.

Inflation erases all inhomogeneities along the flat direction, leaving
only the zero-mode condensate.  The vacuum expectation value (VEV) of
the homogeneous mode can become large during inflation.  In the
post-inflationary epoch the field amplitude is initially damped by the
expansion of the universe and remains essentially fixed.  This stage
ends when the Hubble constant becomes of the order of the flaton mass,
at which point the field starts oscillating around the minimum of the
potential.  This is similar to what happens in chaotic inflation
models, where at the end of inflation the inflaton field starts
oscillating around the minimum of the potential~\cite{linde}.

Reheating of the universe through inflaton decay is a well studied
problem~\cite{Davidson}.  It is now understood that in many models
non-perturbative processes, collectively known as preheating, can lead
to rapid decay of the inflaton
field~\cite{preheating,kofman,cpreheating1,cpreheating2,fpreheating}.
In preheating the decay of the inflaton occurs resonantly, leading to
a rapid amplification of one or more bosons to exponentially large
occupation numbers. This process is eventually halted by the expansion
of the universe, and by the back reaction of the produced quanta. The
decay products scatter of the inflaton field leading to further decay
of the oscillating inflaton zero-mode~\cite{khlebnikov,felder}.

Flat direction condensates with large initial amplitudes may also
undergo resonant decay. There is almost no mention of this possibility
in the existing literature on cosmological scenarios based upon flat
direction condensates. In this paper we analyze the possible
occurrence of flaton preheating.  We start our discussion in the next
section with a review of preheating.  We will highlight the
differences between inflaton and flaton preheating.  We restrict a
detailed discussion of flaton preheating to two specific examples:
Preheating of MSSM flat directions is the subject of section III,
whereas section IV is devoted to resonant decay of the right-handed
sneutrino. We conclude in section V.


\section{Preheating}

Preheating with a real driving field has been discussed in detail in
\cite{preheating,kofman}. Numerical studies of preheating including
the subsequent thermalization phase can be found in
\cite{khlebnikov,felder}. The generalization to complex fields is
described in \cite{cpreheating1,cpreheating2}. We will summarize the
main results.


\subsection{Real driving fields}

The flat direction condensate starts oscillating in its potential when
the Hubble parameter becomes of the order of the flaton mass. Higher
order terms in the potential rapidly become negligible, and the scalar
potential can be written as
\be 
\label{pot}
V(\Phi,\chi) = \half m_\phi^2 \Phi^2 + \half m_\chi^2 \chi^2 +
\half h^2 \Phi^2 \chi^2, 
\ee 
with $\Phi$ the field parameterizing the flat direction, and $\chi$
some other scalar field it couples to.  For the moment we take $\Phi$
real, postponing the discussion of preheating with a complex driving
field to the next subsection. In supersymmetric theories the quartic
coupling can arise from either a Yukawa coupling in the
superpotential, or from gauge interactions.

The equation of motion for the rescaled modes $\chi_k = a^{-3/2}
X_k$ in an expanding FRW universe is
\be \ddot{X}_k + \omega_k^2 X_k = 0\,, 
\ee 
with $k$ the comoving momentum, and a dot denotes differentiation with
respect to time. Further, $\omega_k^2 = {k^2}/{a(t)^2} + m_\chi^2 +
h^2 \Phi^2(t) + \Delta$, with $\Delta = -\frac{3}{4} (\dot{a}/{a})^2 -
\frac{3}{2} (\ddot{a}/a)$.  The scale factor is $a \propto t^{p}$ with
$p = 1/2$ ($p = 2/3$) in a radiation (matter) dominated universe. Then
$\Delta = \frac{3}{2}(\frac{p-1}{p} - \frac{1}{2}) H^2 + \xi R$.  Soon
after the onset of flaton oscillations $\Delta$ becomes negligible
small.  The flaton zero-mode is $\Phi = \phi \cos(m_\phi t)$ with the
amplitude red shifting as $\phi \propto a^{-1/p}$.

The mode equation can be brought in the form of a Mathieu
equation~\cite{mathieu}
\be
X_k'' + ( A_k - 2 q \cos 2z ) X_k = 0\,,
\ee
with $z = m_\phi t$, and prime denotes differentiation with respect to
$z$. Further
\bea
A_k &=& \frac{{k^2/a^2} + m_\chi^2}{m_\phi^2} + 2 q\,, 
\label{A0} \\
q &=& \frac{h^2 \phi^2}{4 m^2}\,.
\label{q0}
\eea 
An important feature of the solutions to the Mathieu equation is the
existence of exponential instabilities. For $q > 1$, many resonance
bands are excited.  Preheating occurs in the {\it broad resonance}
regime where particle production is efficient for modes with momenta
$k^2 \leq A - 2q$.  The occupation numbers of quantum fluctuations
grows exponentially: $n_k \propto \exp(\mu_k z)$ with significant
exponent $\mu_k \sim 0.1$.  The typical momenta $k_*$ of the particles
produced is
%
$k_* = {k_{\rm max}}/{\sqrt{2}} = {m_\phi q^{1/4}}/{\sqrt{2}}$,
%
where we have assumed $m_\chi^2 \ll \sqrt{q} m_\phi^2$.

In an expanding universe preheating attains a stochastic character.
However the net result is still an exponential growth of
$\chi$-fluctuations. Preheating is halted by the expansion of the
universe when $q$ falls below unity.

Effective preheating rests on two principles, violation of
adiabaticity and Bose enhancement. Particle production results from
non-adiabatic changes in the effective frequency of the $X_k$
modes. Adiabaticity is violated when
\be
|\dot{\omega}_k| \gtrsim \omega_k^2, 
\label{non_ad}
\ee
%
%
%
which happens each time the flaton zero-mode goes through the minimum
of the potential and changes rapidly. The occupation numbers of the
decay quanta grow exponentially fast due to Bose-enhancement.  As a
result, preheating is robust.  Resonant production occurs as long as
the non-adiabaticity condition Eq.~(\ref{non_ad}) is fulfilled. Adding
additional fields or couplings (e.g. $h m \Phi \chi^2$) has little or no
effect on the resonant period.  However, the back reaction effects are
very model dependent; when and how they become important depends on
the specifics of the potential.


\subsection{Complex Driving fields}

Supersymmetric theories inherently involve complex scalar fields.
Phase-dependent terms can arise naturally in the potential through
soft SUSY-breaking terms. A relative phase between the oscillations of
the real and imaginary components of the flaton field leads to a
trajectory that is elliptic.  The minimum amplitude is no longer
$|\Phi|_{\rm min} =0$ as it is for a real driving field, but instead
$|\Phi|_{\rm min} =b$ with $b$ the semi-minor axis of the
ellipse. This may prevent adiabaticity violation from occurring.

We can decompose the complex driving field into real and imaginary
components: $\Phi = \Phi_{\rm R} + i \Phi_{\rm I}$.  By a
phase rotation the largest amplitude component of oscillation can be
put in the real piece.  Then
\bea
\Phi_{\rm R} &=& \phi \sin (m_\phi t)\,, \nonumber \\
\Phi_{\rm I} &=& f \phi \cos (m_\phi t)\,,
\eea
with $f = b/a$, the ratio of the semi-minor and semi-major axis
of the elliptic trajectory.  We will refer to $f$ as the
ellipticity of the orbit; note that $f = \sqrt{1-e^2}$ with $e$
the eccentricity of the ellipse. As before, the mode equation for the
$\chi$-quanta in the time-dependent $\Phi$-background can be mapped
into a Mathieu equation, with now~\cite{cpreheating1}
\bea
A_k(f) &=& \frac{{k^2 / a^2} + m_\chi^2 +f^2 h^2 \phi^2}{m_\phi^2}+ 2 q(f)\,,
\\
q(f) &=& (1-f^2) \frac{g^2 \phi^2}{4 m_{\phi}^2}\,. 
\label{q}
\eea 
The $q$-parameter for an elliptic trajectory is reduced by a factor 
$(1-f^2)$, with respect to the case of a pure oscillatory trajectory.  

Adiabaticity violation as defined in Eq.~(\ref{non_ad}) occurs when
\be
\label{effcond}
\frac{k^2}{a^2} + m_\chi^2 + h^2 f^2 \phi^2 + (1-f^2) h^2 \Phi_{\rm R}^2
\lesssim \( (1-f^2) h^2 \Phi_{\rm R} \phi m_\phi \)^{2/3}\,,
\ee
The ellipticity is negligible small when for typical momenta $k \sim
k_*$, the $h^2 f^2 \phi^2$ term in the above equation can be
neglected.  This is the case for
\be
f \lesssim f_{\rm R} \equiv \frac{1}{2 q(0)^{1/4}}\,,
\ee
and preheating proceeds as for a real driving field. For larger
ellipticities the term proportional to $f^2$ in Eq.~(\ref{effcond})
dominates.  This leads to an upper bound on $q(0)$ for which broad
resonance is effective:
\be
q(0) \lesssim q_{\rm c} \equiv \frac{1-f^2}{16f^4}\,.
\label{q_crit}
\ee
The reason is that for larger values of $q(0)$ the semi-minor axis of
the ellipse is large, $b \propto f \sqrt{q(0)}$, and adiabaticity
violation does not occur. For large initial q-values resonant decay is
delayed until the expansion of the universe red shifts the q-parameter
below the critical value.  As discussed below, preheating can be very
efficient if at its onset $q(f) \gtrsim 10^3$, which requires $f
\lesssim 0.1$.  Note that for large ellipticities $f= 1- \epsilon >
0.5$ the upper bound requires $q \lesssim \epsilon$, and preheating
never takes place.

In supersymmetric theories bosonic preheating is generically
accompanied by fermionic preheating, since a Yukawa coupling in the
superpotential leads also to a fermion coupling of the form
\be
{\mathcal L} \ni (m_{\rm f} + h \phi) \bar{\psi}{\psi},
\ee
with $m_{\rm f}$ the fermion mass. Resonant production of fermions has
been studied in~\cite{fpreheating}. As long as the flaton amplitude is
larger than $m_{\rm f}/h$ (i.e., $q \gtrsim 1$), and the ellipticity
is sufficiently small, there is an instant during each flaton
oscillation that the effective mass of the fermion vanishes, and
fermions are produced.  Within about ten oscillations the fermion
occupation number is saturated at a time-averaged value $n_\psi \sim
1/2$ for momenta within a Fermi sphere of radius $k_{\rm f} \sim
m_\phi q^{1/4}$.  The Pauli-exclusion principle forbids occupation
numbers beyond one.  The back reaction of fermions can catalyze
bosonic preheating~\cite{fpreheating}.

Another interesting property of supersymmetric theories with regard to
preheating was observed in~\cite{cpreheating2}, who studied a
superpotential of the form $W = \frac{1}{2} S^2 + \frac{1}{2}(m_\chi +
h \phi) \chi^2$.  After diagonalizing the mass matrix for the real and
imaginary component of $\chi$, it turns out that one of the masses can
become tachyonic during part of the oscillation period of $\phi$. As a
result, quantum fluctuations of $\chi_k$ grow exponentially. We will
refer to this kind of instability as tachyonic preheating.  The
occurrence of tachyonic modes is quite model dependent.

\subsection{Back reaction effects}

Which back reaction effects play an important r\^ole, and at what
stage, is rather model dependent.  We will list here four different
back reaction effects for bosonic preheating.
\begin{itemize}
\item
When the quantum fluctuations of $\chi$ grow exponentially large, the
effective flaton mass $m^2_{\rm eff}=m_\phi^2 + g^2 \langle \chi^2
\rangle$ becomes dominated by the variance term.  The oscillation
frequency of $\Phi$ rapidly increases, and energy is rapidly dumped
into $\chi$-particles until $q \sim 1$ and preheating is halted by the
expansion of the universe.  At the end of this phase $\langle \chi^2
\rangle \sim \phi^2$, and occupation numbers are enormously large $n_k
\sim 10^2 h^{-2}$.  Initial values $q \gtrsim 10^3$ are needed for
this rapid energy transfer to occur; for smaller values preheating is
halted by the expansion of the universe before the variance term comes
to dominate the flaton mass~\cite{kofman}.
\item
The non-zero variance $\langle \chi^2 \rangle$ can also induce an
effective mass for the $\chi$-field itself if quartic couplings
$\lambda \chi^4$ are present in the potential.  Preheating of
particles with masses $m_\chi \gg m_\phi$ is generically inefficient.
This may halt preheating before the stage of rapid energy transfer,
rendering preheating inefficient.
\item
The decay quanta scatter with the zero mode, leading to exponential
amplification of flaton quanta.  However, scattering only becomes
important when the occupation numbers of the $\chi_k$ quanta become
exponentially large, in the very last stages of efficient preheating.
Scattering leads to further decay of the zero-mode.
\item
Flaton preheating can occur in the presence of a thermal bath.  If the
interaction rate for scattering of $\chi$ quanta with the particles in
the thermal bath is much greater than the flaton oscillation
frequency, resonance modes are depopulated rapidly on the relevant
time scale.  There is no Bose enhancement, and resonant production is
inefficient.
\end{itemize}

\subsection{Flaton vs. Inflaton}

In this subsection we consider the differences between resonant decay
of generic flat direction fields and the usual considered case of
resonant decay of the inflaton field.
\begin{itemize}
\item
The energy density stored in the inflaton field dominates the energy
density in the universe during the epoch of preheating.  As a result,
the expansion rate of the universe is set by the inflaton field
itself, and is initially that of matter domination.  The flaton field
on the other hand is generically subdominant at decay. It evolves in a
fixed background, which can be either radiation or matter dominated.
This difference in universe evolution leads to only small changes in
the various quantities, leaving the order of magnitude estimates
unchanged.
\item
The universe is reheated by the decay of the inflaton field.  Thus
inflaton preheating occurs in an empty universe.  This is not
necessarily the case for flaton preheating.  Note in this respect that
even before the reheating process of inflation is completed there is a
dilute plasma with temperature $T \sim (T_R^2 \mpl H)^{1/4}$, where
$T_R$ is the reheat temperature of the universe.  The effects of the
thermal bath should be taken into account for flaton preheating.
\item
The inflaton is usually chosen to be a gauge singlet of the standard
model, its properties largely unconstrained by particle physics or
experimental data.  Moreover, preheating is mostly studied in the
simplest setting: the inflaton is a real field, with only a quartic
interaction term.  Flat direction fields naturally appear in SUSY
theories.  This means that for flaton preheating we can no longer
ignore the complexity of the driving field.  Furthermore, we restrict
our attention to two specific examples, namely the right handed
sneutrino and MSSM flat directions.  For these flatons the potential
is constrained by supersymmetry and particle physics, and there is no
freedom in ignoring/adding unwanted/wanted terms. 
\end{itemize}

\section{MSSM flat directions}

Directions in field space along which the potential vanishes
identically are called flat directions. In supersymmetric theories
there are generically many flat directions at the classical level.
The classical degeneracy along flat directions is protected from
perturbative quantum corrections in the supersymmetric limit by the
non-renormalization theorem. The flatness is lifted by soft terms from
supersymmetry breaking and non-renormalizable operators. During
inflation the inflaton potential dominates the energy density in the
universe. The non-zero energy density breaks supersymmetry leading to
soft masses and $A$-terms for scalar fields~\cite{softmass}.  In this
section we will discuss MSSM flat directions in detail, whereas the
right handed sneutrino field is the subject of the next section.

The MSSM flat directions can be parametrized by gauge invariant
operators, $X$, formed from the product of $m$ chiral superfields
making up the flat direction.  Defining $X = c \phi^m$, the effective
potential for the MSSM flat direction is of the form~\cite{AD,flat}
\be
V = V_m(\Phi) + c_H H^2 |\Phi|^2+ \( \frac{A m_{3/2}+ a H} {n
M^{n-3}} \lambda \Phi^n + {\rm h.c.} \)+ 
\frac{|\lambda|^2 |\Phi|^{2n-2}}{M^{2n-6}}\,.
\ee
Here $M$ is the cutoff scale, typically the GUT or the Planck scale.
In the MSSM with parity conservation, most flat direction are lifted
by $n=4,5$ or $6$ non-renormalizable operators. The flattest one is
lifted by $n=9$. The {\it Hubble induced} terms proportional to $H$
are the soft terms from SUSY breaking by the finite energy density in
the universe.  $V_m$ and the term proportional to the gravitino mass
$m_{3/2}$ are the MSSM soft mass and A-term respectively. The mass
term depends on the SUSY breaking scheme.  For gravity mediated SUSY
breaking $V_m = m^2_{3/2} |\Phi|^2$ at tree level.  In the case of
gauge mediation $V_m = m_\phi^4 \log(1 +|\Phi|^2/m_\phi^2)$, where
$m_{\phi}\sim 1-100$~TeV.

Condensation along the flat direction occurs during inflation, if (1)
the flaton mass is much smaller than the Hubble constant and quantum
fluctuations can lead to large initial field values; or if (2) the
effective flaton mass squared is negative and the field settles at a
minimum away from the origin. The natural size for the soft masses
during inflation is of the order of the Hubble
constant~\cite{AD,softmass}.  Non-minimal K\"ahler potentials can
induce negative soft $({\rm mass})^2$ terms, with $c_H < 0$, realizing
possibility (2). Scenario (1) can be realized in the context of e.g.
$D$-term inflation~\cite{dterm} or no-scale type supergravity
models~\cite{noscale}, where symmetries forbid soft mass terms at tree
level and $c_H,\,a \ll 1$ naturally.

Writing $\Phi= \frac{1}{\sqrt{2}} \phi e^{i \theta}$, the $\theta$
dependent part of the potential reads
\be
V_A  \sim\frac{|\lambda|\phi^n}{M^{n-3}} \( |a|H \cos(\theta_a+ 
\theta_\lambda+ n \theta) 
+|A| m_{3/2} \cos(\theta_A+ \theta_\lambda+ n \theta)\)\,,
\label{A}
\ee
with $\theta_a$, $\theta_A$ and $\theta_\lambda$ the phase of $a$, $A$
and $\lambda$ respectively.  During inflation the $\theta$-field will
settle in one of the minima of the $a$-term if $a \sim {\mathcal
O}(1)$. For $a \ll 1$ the radial motion is dominated by quantum
fluctuations.

In the post-inflationary epoch the evolution of the flaton field is
given by its equations of motion:
%
$\ddot{\phi} + 3 H \dot{\phi} + {\partial V}/{\partial \phi}=0$.
%
The $\phi$ field is damped as long as $H>m_{3/2}$. For negative $({\rm
mass})^2$ ($c_H<0$) the damping is critical, and the field closely
tracks its instantaneous minimum
\be
\phi_{\rm min}(H) \sim \(\frac{\beta H M^{n-3}}{|\lambda|}\)^{1/n-2}\,,
\label{phi_min}
\ee
with $\beta$ some numerical constant depending on $a$, $c_H$ and $n$.
Quantum fluctuations of fields with a positive $({\rm mass})^2$ ($c_H
\geq 0$) saturate at $V(\phi) \sim H^4$~\cite{qf}, and initial
amplitudes $\phi_0 \sim H_I^2/m$ are expected, with $H_I$ the Hubble
constant during inflation.  The field is over damped at the end of
inflation, and remains essentially fixed until $\phi_0 \sim \phi_{\rm
min}$, and it starts slow rolling in the non-perturbative potential.

\begin{figure}[t]
\centering
\hspace*{-5.5mm}
\leavevmode\epsfysize=6cm \epsfbox{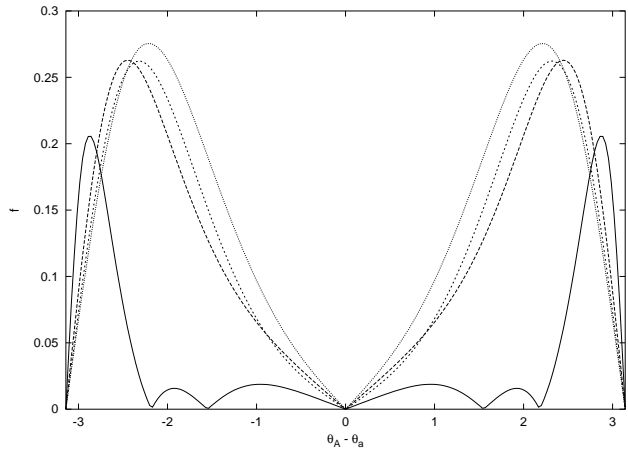}\\
\caption{The ellipticity $f$ as a function of $\theta_A - \theta_a$
for different Hubble induced masses $m^2 = c H^2$.  Solid lines
correspond to $c=-5$, dashed lines to $c=-1$, short dashed lines to
$c=-.25$ and dotted lines to $c=10^{-2}$. All plots is for $n=4$ in a
radiation dominated background.}
\end{figure}

Eventually the Hubble induced mass becomes equal to the soft mass $H
\sim \sqrt{V_m''}$, and the flaton starts to oscillate in the
potential well. This is the moment ellipticity is created.  In gravity
mediated SUSY breaking this phase starts when $H \sim m_{3/2}$. At
this time the Hubble induced and MSSM $A$-terms are of comparable
magnitude and there is a torque in the angular direction if
$\theta_A\neq \theta_a$.  We expect ellipticities of order unity. In
the gauge mediated case $m_{3/2} < \GeV$ and the MSSM $A$-term is
small compared to the Hubble induced one at the onset of
oscillations. Therefore smaller ellipticities are expected. The
ellipticity asymptotes to a constant when $H \sim 0.1 m_{\phi}$.

This is confirmed by numerical calculations~\cite{Jokinen}.  In
gravity mediated SUSY breaking the ellipticity is $f\lesssim 0.5 -
0.2$ for $n=4 - 6$. In more than half of the parameter space
$(\theta_a -\theta_A)\in [0,2\pi]$ the ellipticity is $f < 0.1$. The
parameters chosen in these simulations are $M = \mpl$ and
$|A|=|a|=-|c_H| =1$.  The results are independent of the gravitino
mass and the cutoff scale.  The ellipticities obtained in the gauge
mediated case are smaller $f \lesssim 0.1 - 0.01 $ for $n=4-6$.
These results are for $m_{3/2} = 10^{-5}, 10^{-9}m_\phi$; for much
larger values the behavior as in the gravity mediated case.

\begin{figure}[t!]
\centering
\hspace*{-5.5mm} 
\leavevmode\epsfysize=6cm \epsfbox{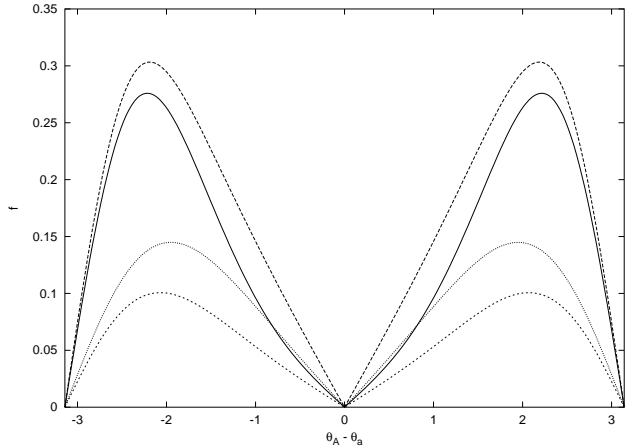}\\
\caption{Ellipticity for $F$-term inflation $a\neq 0$ and $n=4$ (solid
lines); $D$-term inflation $a=0$ and $n=4$ (dashed lines); radiation
domination and $n=6$ (short dashed lines); and for matter domination
and $n=6$ (dotted lines). For all plots $c=0$.}
\end{figure}

We have extended the numerical calculations of~\cite{Jokinen} to study
the elliptical trajectory for more general parameters, focusing on a
mass term of the form $V_m = m_\phi^2 |\Phi|^2$. We found that the
ellipticity has the same order of magnitude independent of the sign of
the Hubble induced masses, i.e., it is independent of $c_H$, as shown in
Fig.~1. Further, the same results are obtained in both $F$-term and
$D$-term inflation, see Fig.~2. In the latter no Hubble induced terms
are generated, and $a=0$. 

Ellipticities do depend sensitively on the ratio of the gravitino to
flaton mass, and also on the initial amplitude of the flaton
field. The ellipticity is suppressed by a factor
$m_{3/2}/m_\phi$. This is as expected, as in the limit $m_{3/2}/m_\phi
\to 0$ no potential in the angular direction is generated.  This is
shown in Fig.~3. For negative $({\rm mass})^2$ the field is trapped in
the minimum of the potential, given by Eq.~(\ref{phi_min}).  But for
positive $({\rm mass})^2$ it is possible to have an initial amplitude
differing from $ \phi_{\rm min}$. Ellipticities decrease rapidly for
$\phi_0 < \phi_{\rm min}$, as shown in Fig.~4. For the parameter
values $m_{3/2} \sim \TeV$ and $M = \mpl$, $\phi_{\rm min} = 10^{11} -
10^{14} \GeV$ for $n = 4 -6$ at the onset of oscillations.  Quantum
fluctuations saturate at a field value lower than $\phi_{\rm min}$ for
Hubble constants during inflation $H_I \lesssim 10^7 - 10^9 \GeV$ for
$n =4 -6$.

\begin{figure}[t!]
\centering
\hspace*{-5.5mm}
\leavevmode\epsfysize=6cm \epsfbox{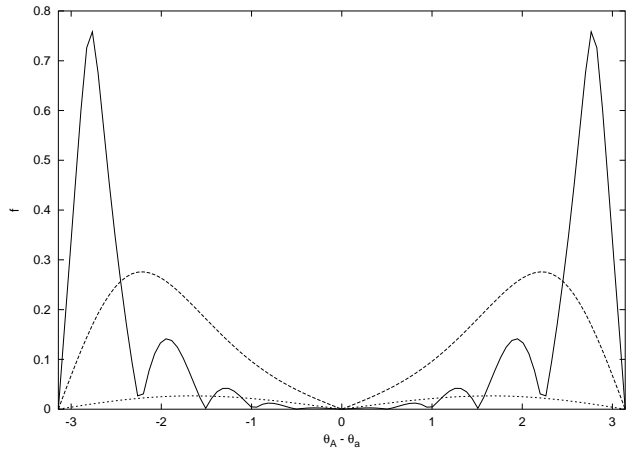}\\
\caption{Ellipticity for different ratios of the gravitino to flaton
mass.  Solid lines correspond to $m_{3/2}/m = 3$, dashed lines to
$m_{3/2}/m = 1$, and dotted lines to $m_{3/2}/m = 0.1$. For all plots $c=0$.}
\end{figure}

The flaton has $D$-term couplings to the fields making up the
polynomial $X$, which are orhtogonal to the direction which aquires
the VEV~\cite{AD_thermal}.  In addition, the flaton has $F$-term
couplings to fields not present in $X$.  The interaction term is of
the form $h^2 \phi^2 \chi^2$, with $h$ either a gauge or yukawa
coupling.  The $q$-parameter is $q(f)=(1-f^2)h^2\phi^2/(4
m_\chi^2)$. The gauge and Yukawa couplings in the MSSM range between
$h \sim 1-10^{-6}$, $m_\chi = m_{3/2} \sim 1\TeV$, and the field
amplitude at $H\sim m_\phi$ is given by Eq.~(\ref{phi_min}). Typically
$q(0)\gg 1$. We can then distinguish the following possibilities.

\begin{enumerate}

\item
The ellipticity is negligible small. Preheating starts when $H\sim
m_\phi$ and can be effective, depending on the back reaction
effects. Such small ellipticities are obtained for fine-tuned
parameters $\theta_A \approx \theta_a$, for $\phi_0 \ll \phi_{\rm
min}$ as is possible for positive (mass)$^2$ and low scale inflation,
or in the absence of $A$-terms.  Generically, supergravity corrections
will induce non-zero $A$-terms.

\item
The ellipticity is small $f \lesssim 0.1$. This is the case in gauge
mediated SUSY breaking, and also in most of the parameter space for
gravity mediation. Moreover, ellipticities are suppressed if
$m_{3/2}/m_\phi <1$ and/or $\phi_0/\phi_{\rm min} <1$.  Preheating
starts off in the broad-band regime. However, its onset is delayed
until $q(0)\lesssim q_{\rm c}\approx 1/(16 f^4)$, and therefore might
not occur if the condensate decays perturbatively, via thermal
scattering or through fragmentation into Q-balls before this
time. Note that the decay quanta that will be first exited are the
fields corresponding to the smallest $q$-parameter, i.e., the fields
with the smallest coupling to the flaton.

Denote the decay width of the condensate by $\Gamma_\phi \sim \beta
m_\phi$.  For perturbative decay $\beta = h^2$ with $h$ a Yukawa or
gauge coupling.  For temperatures higher than the effective mass of
the particles coupling to the flaton, i.e., $T \gtrsim (m_\chi)_{\rm
eff} \sim h \phi$, there is a thermal bath of $\chi$-particles.
Thermal scattering can lead to decay of the condensate; in this case
$\beta \sim h^2 \alpha (T/m_\phi)$ with $\alpha =g^2/4\pi$.  Note that
in the presence of a thermal bath the flaton mass should be corrected
by the thermal contribution in all formulas $\delta m^2_\phi \sim h^2
T^2$.  If it dominates, the thermal mass induces early oscillations.
Now the field amplitude, and therefore $q$, not only decreases because
of the red shift, but also because the effective mass decreases.  This
will speed up the onset of preheating, since $q(0) < q_c$
earlier. Thermal effects are particulary important for $n=4$
directions with small Yukawa couplings $h \lesssim
10^{-3}$~\cite{AD_thermal}. Since the thermal history is rather model
dependent, we will not pursue this issue further.  Finally, if the
flaton potential grows less than $\phi^2$ the condensate is unstable
against fragmentation into $Q$-balls. Numerical simulations indicate
that $Q$-ball formation takes place at $H\sim 10^{-4}-10^{-6}m_\phi$
in gravity and gauge mediated scenarios respectively~\cite{kasuya}.

If we denote the Hubble constant at the moment $q(0)=q_{\rm c}$ with
$H_c$, then preheating can play a r\^ole if $H_c \gtrsim
\Gamma_\phi$. In a matter dominated universe this translates into
$\alpha^{3/4} f^2 h \lesssim m_\phi / \phi_{\rm min}$, with $\phi_{\rm
min}$ evaluated at $H \sim m_\phi$.  The universe is matter
(radiation) dominated before (after) inflaton decay.  For example, for
$m_\phi \sim \TeV$ and $M \sim \mpl$ preheating occurs before
perturbative decay if $h^{5/2} f^2 < 10^{-8},\, (10^{-12})$ for $n=4,
\,(6)$.  Preheating precedes $Q$-ball formation if $f^2 h \lesssim
10^{-6} - 10^{-9}$ for $n=4-6$ and gravity mediated SUSY breaking, and
$f^2 h \lesssim 10^{-4} - 10^{-8}$ for $n=4-6$ and gauge mediated SUSY
breaking.

\item
 The ellipticity is appreciable $0.2 < f < 1$.  Preheating is
 delayed. If it takes place at all, it is shut off by the expansion of
 the universe before the back reaction becomes important. Large, order
 one, ellipticities are possible in gravity mediated SUSY breaking
 with $n=4$ non-renormalizable operators.

\end{enumerate}

\begin{figure}[t]
\centering
\hspace*{-5.5mm}
\leavevmode\epsfysize=6cm \epsfbox{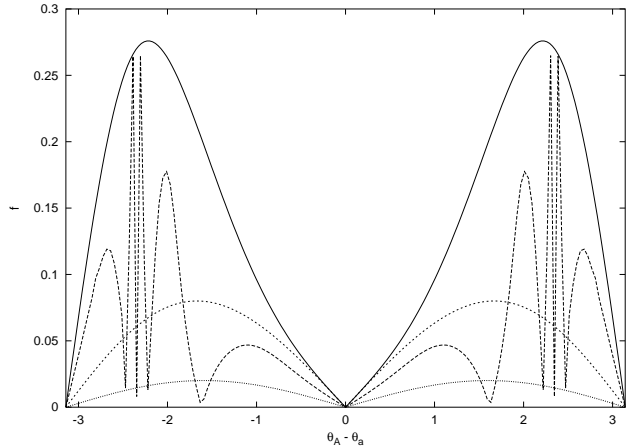}\\
\caption{Ellipticity for different values of the initial amplitude.
Solid lines correspond to $\phi_0/\phi_{\rm min} = 1$, dashed lines to
$\phi_0/\phi_{\rm min} = 5$, short dashed lines to $\phi_0/\phi_{\rm
min} = 0.1$, and dotted lines to $\phi_0/\phi_{\rm min} = 0.05$. For
all plots $c=0$.}
\end{figure}

\subsection{backreaction}

Quartic self-couplings of the decay mode $V \ni h^2 \chi^4$ will
induce an effective mass term $m_\chi \sim h^2 \langle \chi^2 \rangle$
as the variance grows.  All MSSM bosons (except for the right handed
sneutrino) are charged under the standard model gauge group, and have
a quartic interaction with $h$ the gauge coupling.  In addition, the
superpotential term $W \ni h \phi \chi^2$ leads to a quartic
interaction in the $F$-term potential. It has been found that for a
potential of the form $V(\phi) + m_\chi^2 \chi^2 + h^2 \phi^2 \chi^2$
resonant production is only effective for couplings $h^2 \gtrsim
10^{-7} (m_\chi/m_\phi)^4$~\cite{khlebnikov}. Therefore, bosonic
preheating shuts off when $(m_\chi)_{\rm eff} \sim m_\phi$.  From
$n_\chi = g \phi_0 \langle \chi^2 \rangle$~\cite{kofman}, it follows
$\rho_\chi \sim m_\phi^3 \phi_0 /h \ll \rho_\phi \sim m_\phi^2
\phi_0^2$.  Thus at the of preheating, only a small fraction of the
energy stored in the flaton zero mode will be transferred to the decay
products.

No effective mass term for the fermion superpartners of $\chi$ are
generated. However, the Pauli exclusion principle forbids occupation
numbers exceeding unity.  The typical energy transferred to the
fermions is $k_F^4 \sim h^2 m_\phi^2 \phi_0^2$, which is only
significant for large couplings.

The typical interaction rate for MSSM particles in a thermal bath is
of the order $\sigma \sim \alpha^2 T$. The relevant time scale in
preheating is $m_\phi^{-1}$.  It follows that thermal scattering is
important for $\alpha^2 \gtrsim m_\phi/ \sqrt{H \mpl}$ for $T < T_R$,
and $\alpha^2 \gtrsim m_\phi/ (T_R^2 H \mpl)^{1/4}$ for $T > T_R$,
with $T_R$ the reheat temperature of the universe.  Effective
scattering kills Bose-Enhancement, but can increase the efficiency of
fermionic preheating, since in between each moment of particle
production the decay products are scattered out of the resonance
bands.

\section{Right handed sneutrino field}

The right-handed neutrino and its scalar partner appear in grand
unified theories (GUT) based upon $SO(10)$. At first sight it appears
the sneutrino direction is lifted by a $D$-term potential due to the
$SO(10)$ gauge interactions, preventing a large VEV for the sneutrino
field $\tilde{N}$. The crucial observation made in~\cite{sneutrino}
however is that the $D$-term decouples from the potential of
$\tilde{N}$ as long as the value of $\tilde{N}$ is smaller than the
$SO(10)$ breaking scale. The sneutrino field condenses during
inflation if $m^2_N \ll H^2$, with a maximum amplitude set
by the GUT scale.

The superpotential containing the three right-handed sneutrino fields
(after $B-L$ breaking) is
\be
W=\sum_{i=1}^3\half (m_N)_i N_i N_i +\sum_{i,\alpha = 1}^3 h_{i\alpha} 
N_i L_\alpha H\,,
\ee
here $N_i$ are the three sneutrino superfields, $L_\alpha$ the three
left-handed lepton doublets, $H$ is shorthand for $H_u$ the down Higgs
superfield, and $h_{i\alpha}$ are the Yukawa couplings.  We start our
discussion of preheating by considering a single generation.  Writing
the scalar potential in terms of the component fields $L^T =( L^0 \:
L^-)$ and $H^T =( H^+ \: H^0)$
\bea 
V_F &=& m_N^2 N^2 + |h|^2 |N|^2 (|L|^2 + |H|^2) + h^* m_N N
(L^0 H^0 - L^- H^+)^*  + c.c. \\ 
V_D &=& \frac{g_z^2}{8} (|L^0|^2 - |H^0|^2)^2
+ \frac{g_W^2}{2}  (L^{0*} L^- + H^{+*} H^0)^2 + c.c,\\
V_S  &=& m_L |L|^2 + m_H |H|^2  + (B m_{3/2}^2 + b H^2) N^2 
+ \frac{A m_{3/2} +a H}{ n M^{n-3}} N^n + c.c.
\label{V_N}
\eea
with $|L|^2 \equiv |L^0|^2 + |L^-|^2$ and $|H|^2 \equiv |H^+|^2 +
|H^0|^2$. $V_F$ and $ V_D$ are the $F$ and $D$-term potential
respectively, and $V_S$ is the soft SUSY breaking potential. The
coefficients $B,\,b,\,A\,,a \sim {\mathcal O}(1)$, $g_W = e / \sin
\theta_W$ and $g_Z = e / \sin \theta_W \cos \theta_W$. The $a,
\,b$-terms are from SUSY breaking by the finite energy density in the
universe, whereas the $A, \, B$-terms arise from low energy SUSY
breaking. $\Lambda$ is the ultraviolet cutoff scale, typically
$\Lambda \sim M_{\rm GUT}$ or $\Lambda \sim \mpl$.  If R-parity is
conserved the lowest order $A$-term for the sneutrino has $n=4$.
Furthermore, we have assumed that non-renormalizable operators are
sub-dominant.  An inclusion of these operators will not affect the
results in any essential way.

Since the $B$-terms do not contain a coupling between the real and
imaginary parts of $N$, it will not induce an ellipticity.  The sole
source of ellipticity are the $A$-terms.  During inflation the Hubble
induced $A$-term is dominant and the the $\theta$ field quickly
settles in one of the minima of $\cos(\theta_A+\theta_\lambda
+n\theta)$. At the onset of oscillations, when $H\sim m_N$, the low
energy $A$-term is negligible small for gravitino masses $m_{3/2}\ll
m_N$, and the field remains essentially fixed in its $\theta$-minimum.
The field amplitude red shift with time and all $A$-terms quickly
become sub-dominant; hereafter no further ellipticity is generated.
We expect therefore that for sneutrino masses $m_N \gg m_{3/2}$ the
ellipticity is negligible small, and we can use the theory of
preheating for a real driving field. This is confirmed by our
numerical calculations which give $f\sim 0.1(m_{3/2}/ m_N)$, see
Fig.~4.  Therefore we take $N$ real.  For now we also take the yukawa
coupling $h$ real, deferring a discussion of a complex coupling to
section~\ref{section_cp}.

We decompose the slepton and Higgs fields in its real and imaginary
component. Diagonalizing the mass matrix gives the following mass
eigenvalues and eigenstates:
\bea
M^2_{R\pm} &= h N( h N \pm m_N), 
&\quad  {\rm for} \; X_{R\pm} = L^0_R \pm H^0_R, \\
M^2_{I\pm} &= h N( h N \pm m_N),  
&\quad  {\rm for} \; X_{I\pm} = L^0_I \mp H^0_I, \\
\tilde{M}^2_{R\pm} &= h N( h N \pm m_N), 
&\quad  {\rm for} \; \tilde{X}_{R\pm} = L^-_R \mp H^+_R, \\
\tilde{M}^2_{I\pm} &= h N( h N \pm m_N),  
&\quad  {\rm for} \; \tilde{X}_{I\pm} = L^-_I \pm H^+_I,
\eea
where the unbarred (barred) quantities correspond to the neutral
(charged) fields.  

Due to both the linear and quadratic term in $N$ we cannot map the
mode equation for the decay products in terms of a Mathieu
equation. However, adiabaticity violation and thus particle production
occurs when $N$ goes through zero, and
\be
| \dot{\omega}_k| \gtrsim \omega_k^2.
\ee
During most of the resonance time the quadratic term dominates over
the linear term.  Hence, the physics is well captured by this term
alone, and the problem can be rewritten in terms of a Mathieu equation
with $q= h^2 N_0^2/(4 m_N^2)$.  Particle production occurs in a time
interval $\delta t_{\rm non-ad} \sim (h m_N N_0)^{-1/2}$.

The (mass)$^2$ eigenvalues are not positive definite.  During part of
the $N$-oscillation the masses $M_{-}^2$ can go negative, and quantum
fluctuations of the $X_{-}$ fields grow exponentially due to the
tachyonic instability.  For $h N_0 \lesssim m_n$ or equivalently $q
\lesssim 1$, the various mass terms become tachyonic during half of
the sneutrino oscillation, and tachyonic preheating is very efficient.
On the other hand, for $q \gtrsim 1$ the mass terms only become
tachyonic during the small time interval $\delta t_{\rm tach} \sim 1
/(h N_0)$. Since $(\delta t)_{\rm non-ad} \gg (\delta t)_{\rm tach}$,
it is expected that in this regime tachyonic particle production is
sub-dominant.  Particles with masses up to $m_N^2/4$ can be produced
during tachyonic preheating. .

The $q$-parameter for sneutrino decay is $q^2 = h^2 N_0^2 / 4 m_N^2$.
In the seesaw mechanism the neutrino masses are related to the Yukawa
couplings and sneutrino masses through
$ m_\nu = \frac{h^2 \langle H_u \rangle^2}{m_N}$,
where $\langle H_u\rangle= 174\GeV\times\sin\beta$ and
$\tan\beta=\langle H_u\rangle/\langle H_d\rangle$.  The solar and
atmospheric neutrinos have masses in the range $10^{-1}-10^{-3}\eV$.
Typical values of the q-parameter are
\be 
 q \sim 10^{5}
\(\frac{m_\nu}{10^{-2} \eV}\)
\(\frac{10^{12} \GeV}{m_N}\)
\(\frac{N_0}{10^{16} \GeV}\)^2\,.
\ee
Typically $q\gg 1$, and we expect non-adiabatic preheating to happen
generically.  Note however that there is no lower bound on the
lightest neutrino mass, and therefore some of the Yukawa couplings
might be arbitrarily small.  Moreover it might be that $N$ gives a
negligible contribution to the neutrino mass, which are dominated by
their couplings to the the sneutrinos in the other two families. In
those cases the $q$-parameter can be lowered to arbitrary low values.

\subsection{Back reaction effects}

The quartic interaction terms can be written in terms of the mass
eigenstates:
\bea
V_D &= &\frac{g_z^2}{8} (X_{I+} X_{I-} + X_{R+} X_{R-})^2
-\frac{g_w^2}{4} 
(X_{R+} \tilde{X}_{I-} + X_{R-}\tilde{X}_{I+} -
X_{I+} \tilde{X}_{R-} - X_{I-}\tilde{X}_{R+} )^2
\nonumber \\
&&+\frac{g_w^2}{4} 
(X_{I+} \tilde{X}_{I-} + X_{I-}\tilde{X}_{I+} +
X_{R+} \tilde{X}_{R-} + X_{R-}\tilde{X}_{R+} )^2
\\
V_F &\ni& \frac{h^2}{16} \( (X_{2-} -X_{2+})^2 + (X_{1+} -X_{1-})^2 \) 
\( (X_{2-} + X_{2+})^2 + (X_{1+} + X_{1-})^2 \) \nonumber \\
\eea
Consider first the case $q \gg 1$.  Then tachyonic instabilities are
ineffective, and the eigenstates $X_i$ all grow with the same
rate. The $D$-term potential generates effective mass terms for all
modes $(m_X)_{\rm eff} \sim g^2 \langle X^2 \rangle$.  Just as in the
MSSM case, this effective mass will halt preheating before a
significant amount of energy is transferred from the sneutrino zero
mode to the sleptons and Higgses.  This conclusion can be avoided if
one of the $X$-fields has a large VEV before the start of preheating.
For example, if $\langle X_{I+}\rangle$ is initially large, it gives a
large mass to the $X_{I-}$ and $\tilde{X}_{I-}$ fields.  As a result,
these eigenstates are not exited during preheating, and no effective
mass for the $X_{I+}$ field is generated.

The case $q \lesssim 1$ is quite different.  Tachyonic preheating is
effective, and only the tachyonic eigenstates $X_-$ are produced.  The
$D$-term will not generate an effective mass state for these fields.
The $F$-term potential does lead to an effective mass term $m_{X_-}
\sim h^2 \langle X_-^2 \rangle$.  Preheating halts when the effective
mass term becomes of the order of the right handed sneutrino mass.
Before this time a significant amount of energy can be transferred
from the sneutrino zero mode to the tachyonic fields.

The last back reaction effect to be considered is thermal scattering.
The typical interaction rate for Higgs, selectron and left handed
sneutrino particles in a thermal bath is of the order $\Gamma \sim T$,
$\Gamma \sim \alpha^2 T$ and $\Gamma \sim G_F^2 T^5$ respectively,
with $G_F$ the fermi constant. The relevant time scale in preheating
is $m_\phi^{-1}$.  It follows that scattering is important, both for
the Higgs and slepton fields, even before inflaton decay when there is
a dilute plasma with temperature $T \sim (T_R^2 H \mpl)^{1/4}$. Decay
quanta are scattered rapidly out of the resonance bands. It requires
further studies to determine how effective thermal scattering is in
killing the resononance.  The reason is that tachyonic preheating can
be very rapidly itself: In the absence of a thermal bath the zero mode
typically decays within one or two oscillations~\cite{tachyonic}.

\subsection{CP violation}
\label{section_cp}

To study CP violation during preheating we allow the coupling constant
$h$ to be complex.  Particle production due to violation of
non-adiabaticity is dominated by the quartic CP conserving term $h^2
N^2 \chi^2$. Hence we expect CP violation to be small.

To study CP violation for the tachyonic decay mode, we decompose $h =
h_R + i h_I$ and diagonalize the mass matrix.  The mass eigenvalues
and eigenstates are:
\bea
M^2_{R\pm} &= |h| N( |h| N \pm M)  
&\quad  {\rm for} \; 
X_{R\pm} = L^0_R \pm \frac{h_R}{|h|}H^0_R \mp \frac{h_I}{|h|}H^0_I \\
M^2_{I\pm} &= |h| N( |h| N \pm M)  
&\quad  {\rm for} \; 
X_{I\pm} = L^0_I \mp \frac{h_R}{|h|} H^0_I \mp \frac{h_I}{|h|}H^0_R \\
\eea
Similar expression holds for the charged lepton and charged higgs
fields.  It follows that tachyonic preheating will not generate a
lepton asymmetry.

Finally we consider the fermionic decay modes.  The mass matrix for
the neutrino and higgsino, in the basis $(\nu_L \:\; \tilde{H}^0)$, is
of the form:
\be
\begin{pmatrix}
0 & h N \\
h^* N & 0
\end{pmatrix}.
\ee 
Although the fermion mass matrix is hermitian, CP is broken since for
a complex coupling since $M \neq M^*$.  A lepton asymmetry is expected
in fermionic preheating, as discussed in reference ~\cite{coherent}.

\section{Discussion}

We studied non-perturbative, resonant decay of flat direction fields,
concentrating on MSSM flat directions and the right handed
sneutrino. The difference between inflaton preheating and flaton
preheating, is that the potential is more constraint in the latter
case. The effects of a complex driving field, and of quartic couplings
in the potential are important and cannot be neglected.  Moreover,
preheating occurs in the presence of a thermal bath.  Effective
scattering can kill Bose-enhancement effects.

Ellipticities of MSSM flat directions condensate are generically of
order one, thereby delaying the onset of preheating.  Preheating may
be preceded by perturbative or thermal decay, or by the formation of
$Q$-balls. If preheating does occur it is generically ineffective due
to the quartic self-couplings of the decay modes, and due to the
presence of the thermal bath.  

The ellipticity for the right handed sneutrino field is negligible
small. Particle production due to the violation of adiabaticity is
generically expected to occur.  However, just as in the MSSM case, it
is expected to be inefficient due to back reaction effects. The new
feature in the sneutrino potential is that for small values $q = h^2
N_0^2 /(4 m_N) \lesssim 1$, there are tachyonic instabilities.  The
$D$-term quartic couplings do not generate an effective mass for the
tachyonic modes, making it an efficient decay channel.  It is unclear
how thermal scattering affects the resonance.


\section*{Acknowledgments}
A. M. is a Cita National fellow, and M. P. is supported by the
European Union under the RTN contract HPRN-CT-2000-00152 Supersymmetry
in the Early Universe. The authors are thankful to Robert
Brandenberger and Patrick Greene for helpful discussions.


\vskip2pc


\begin{thebibliography}{99}


\bibitem{kt}
E.~Kolb and M.~Turner, {\it The Early Universe}
(Addison-Wesley, 1990).

\bibitem{AD}
I.~Affleck and M.~Dine,
Nucl.\ Phys.\ B {\bf 249}, 361 (1985);
M.~Dine, L.~Randall and S.~Thomas,
Nucl.\ Phys.\ B {\bf 458}, 291 (1996),
[arXiv:hep-ph/9507453].

\bibitem{sneutrino}
H.~Murayama and T.~Yanagida,
Phys.\ Lett.\ B {\bf 322}, 349 (1994),
[arXiv:hep-ph/9310297].
K.~Hamaguchi, H.~Murayama and T.~Yanagida,
Phys.\ Rev.\ D {\bf 65}, 043512 (2002),
[arXiv:hep-ph/0109030].
Z.~Berezhiani, A.~Mazumdar and A.~Perez-Lorenzana,
Phys.\ Lett.\ B {\bf 518}, 282 (2001)
[arXiv:hep-ph/0107239].

\bibitem{qballs}
A.~Kusenko,
Phys.\ Lett.\ B {\bf 405}, 108 (1997)
[arXiv:hep-ph/9704273].
K.~Enqvist and J.~McDonald,
Phys.\ Lett.\ B {\bf 425}, 309 (1998)
[arXiv:hep-ph/9711514].

\bibitem{kasuya}
S.~Kasuya and M.~Kawasaki,
Phys.\ Rev.\ D {\bf 61}, 041301 (2000)
[arXiv:hep-ph/9909509].
S.~Kasuya and M.~Kawasaki,
Phys.\ Rev.\ D {\bf 62}, 023512 (2000)
[arXiv:hep-ph/0002285].
{\it for the formation of $Q$-balls without A-terms, see}
K.~Enqvist, S.~Kasuya and A.~Mazumdar,
Phys.\ Rev.\ Lett.\  {\bf 89}, 091301 (2002)
[arXiv:hep-ph/0204270].
K.~Enqvist, S.~Kasuya and A.~Mazumdar,
Phys.\ Rev.\ D {\bf 66}, 043505 (2002)
[arXiv:hep-ph/0206272].


\bibitem{early}
S.~Mollerach,
Phys.\ Rev.\ D {\bf 42}, 313 (1990);
A.~D.~Linde and V.~Mukhanov,
Phys.\ Rev.\ D {\bf 56}, 535 (1997).
[arXiv:astro-ph/9610219];


\bibitem{sloth}
K.~Enqvist and M.~S.~Sloth,
Nucl.\ Phys.\ B {\bf 626}, 395 (2002)
[arXiv:hep-ph/0109214].
T.~Moroi and T.~Takahashi,
Phys.\ Lett.\ B {\bf 522}, 215 (2001)
[Erratum-ibid.\ B {\bf 539}, 303 (2002)]
[arXiv:hep-ph/0110096].
D.~H.~Lyth and D.~Wands,
Phys.\ Lett.\ B {\bf 524}, 5 (2002),
[arXiv:hep-ph/0110002];

\bibitem{curvaton}
K.~Enqvist, S.~Kasuya and A.~Mazumdar,
Phys.\ Rev.\ Lett.\  {\bf 90}, 091302 (2003)
[arXiv:hep-ph/0211147].
K.~Enqvist, A.~Jokinen, S.~Kasuya and A.~Mazumdar,
arXiv:hep-ph/0303165.
M.~Postma,
arXiv:hep-ph/0212005.
D.~H.~Lyth, C.~Ungarelli and D.~Wands,
arXiv:astro-ph/0208055.
N.~Bartolo and A.~R.~Liddle,
Phys.\ Rev.\ D {\bf 65}, 121301 (2002),
[arXiv:astro-ph/0203076];
K.~Dimopoulos and D.~H.~Lyth,
arXiv:hep-ph/0209180];
T.~Moroi and H.~Murayama,
arXiv:hep-ph/0211019;
T.~Moroi and T.~Takahashi,
Phys.\ Rev.\ D {\bf 66}, 063501 (2002)
[arXiv:hep-ph/0206026].






\bibitem{grisaru}
M. Grisaru, W. Sigl, and M. Rocek, Nucl. Phys. B {\bf 159}, 429
(1975); N. Seiberg, Phys. Lett. B {\bf 318}, 469 (1993).

\bibitem{gherghetta}
T.~Gherghetta, C.~F.~Kolda and S.~P.~Martin,
Nucl.\ Phys.\ B {\bf 468}, 37 (1996)
[arXiv:hep-ph/9510370].

\bibitem{flat} 
For a review on MSSM flat directions see by K.~Enqvist and A.~Mazumdar,
arXiv:hep-ph/0209244.
M.~Dine and A.~Kusenko,
arXiv:hep-ph/0303065.



\bibitem{linde}
For a review, see: A.D. Linde, {\it Particle Physics And Inflationary
Cosmology}, Harwood (1990).



\bibitem{Davidson}
S. Davidson, and S. Sarkar, JHEP {\bf 0011}, 012 (2000).
R. Allahverdi, and M. Drees, Phys. Rev. D {\bf 66}, 063513 (2002).
P.~Jaikumar and A.~Mazumdar,
arXiv:hep-ph/0212265.


\bibitem{preheating}
J. Traschen, R. Brandenberger, Phys. Rev. D {\bf 42 }, 2491 (1990);
Y. Shtanov, Ukr. Fiz. Zh. {\bf 38}, 1425 (1993). (in Russian);
Y. Shtanov, J. Traschen, and R. Brandenberger, Phys. Rev. D {\bf 51},
5438 (1995); 
L.~Kofman, A.~D.~Linde and A.~A.~Starobinsky,
Phys.\ Rev.\ Lett.\  {\bf 73}, 3195 (1994)
[arXiv:hep-th/9405187].
P.~B.~Greene, L.~Kofman, A.~D.~Linde and A.~A.~Starobinsky,
Phys.\ Rev.\ D {\bf 56}, 6175 (1997)
[arXiv:hep-ph/9705347].
D.~Boyanovsky, H.~J.~de Vega and R.~Holman,
Phys.\ Rev.\ D {\bf 49}, 2769 (1994)
[arXiv:hep-ph/9310319].
D.~Boyanovsky, H.~J.~de Vega, R.~Holman, D.~S.~Lee and A.~Singh,
Phys.\ Rev.\ D {\bf 51}, 4419 (1995)
[arXiv:hep-ph/9408214].
D.~Boyanovsky, M.~D'Attanasio, H.~J.~de Vega, R.~Holman and D.~S.~Lee,
Phys.\ Rev.\ D {\bf 52}, 6805 (1995)
[arXiv:hep-ph/9507414].
D.~Boyanovsky, H.~J.~de Vega, R.~Holman and J.~F.~Salgado,
Phys.\ Rev.\ D {\bf 54}, 7570 (1996)
[arXiv:hep-ph/9608205].
D.~Boyanovsky, D.~Cormier, H.~J.~de Vega, R.~Holman, A.~Singh and M.~Srednicki,
Phys.\ Rev.\ D {\bf 56}, 1939 (1997)
[arXiv:hep-ph/9703327].
D.~Cormier, K.~Heitmann and A.~Mazumdar,
Phys.\ Rev.\ D {\bf 65}, 083521 (2002)
[arXiv:hep-ph/0105236].

\bibitem{kofman}
L.~Kofman, A.~D.~Linde and A.~A.~Starobinsky,
Phys.\ Rev.\ D {\bf 56}, 3258 (1997)
[arXiv:hep-ph/9704452].

\bibitem{cpreheating1} 
R.~Allahverdi, R.~H.~Shaw and B.~A.~Campbell,
Phys.\ Lett.\ B {\bf 473}, 246 (2000)
[arXiv:hep-ph/9909256].

\bibitem{cpreheating2} 
Z.~Chacko, H.~Murayama and M.~Perelstein,
arXiv:hep-ph/0211369.

\bibitem{fpreheating}
A. D. Dolgov, and D. P. Kirilova, 
Sov.\ J.\ Nucl.\ Phys.\  {\bf 51}, 172 (1990)
[Yad.\ Fiz.\  {\bf 51}, 273 (1990)].
J.~Baacke, K.~Heitmann and C.~Patzold,
Phys.\ Rev.\ D {\bf 58}, 125013 (1998)
[arXiv:hep-ph/9806205].
P.~B.~Greene and L.~Kofman,
Phys.\ Rev.\ D {\bf 62}, 123516 (2000)
[arXiv:hep-ph/0003018].
P.~B.~Greene and L.~Kofman,
Phys.\ Lett.\ B {\bf 448}, 6 (1999)
[arXiv:hep-ph/9807339].
M. Peloso, and L. Sorbo, JHEP {\bf 0005}, 016 (2000);
[arXiv:hep-ph/0003045];
{\it for higher spin fermions, see} 
A. L. Maroto, and A. Mazumdar, Phys. Rev. Lett. {\bf 84}, 1655 (2000)
[arXiv:hep-ph/9904206],




\bibitem{khlebnikov}
S.~Y.~Khlebnikov and I.~I.~Tkachev,
Phys.\ Rev.\ Lett.\  {\bf 77}, 219 (1996)
[arXiv:hep-ph/9603378].
S.~Y.~Khlebnikov and I.~I.~Tkachev,
Phys.\ Lett.\ B {\bf 390}, 80 (1997)
[arXiv:hep-ph/9608458].
S.~Y.~Khlebnikov and I.~I.~Tkachev,
Phys.\ Rev.\ Lett.\  {\bf 79}, 1607 (1997)
[arXiv:hep-ph/9610477].
T.~Prokopec and T.~G.~Roos,
Phys.\ Rev.\ D {\bf 55}, 3768 (1997)
[arXiv:hep-ph/9610400].

\bibitem{felder}
G.~N.~Felder and L.~Kofman,
Phys.\ Rev.\ D {\bf 63}, 103503 (2001)
[arXiv:hep-ph/0011160].

\bibitem{mathieu}
N. W. McLachlan, {\it Theory and Applications of Mathieu Functions}, Dover Publications,
New York (1961).


\bibitem{softmass}
M.~Dine, W.~Fischler and D.~Nemeschansky,
Phys.\ Lett.\ B {\bf 136}, 169 (1984);
O.~Bertolami and G.~G.~Ross,
Phys.\ Lett.\ B {\bf 183}, 163 (1987);



\bibitem{dterm}
P.~Binetruy and G.~R.~Dvali,
Phys.\ Lett.\ B {\bf 388}, 241 (1996),
[arXiv:hep-ph/9606342];
C.~F.~Kolda and J.~March-Russell,
Phys.\ Rev.\ D {\bf 60}, 023504 (1999),
[arXiv:hep-ph/9802358].

\bibitem{noscale}
M.~K.~Gaillard, H.~Murayama and K.~A.~Olive,
Phys.\ Lett.\ B {\bf 355}, 71 (1995),
[arXiv:hep-ph/9504307].


\bibitem{qf}
T.~S.~Bunch and P.~C.~Davies,
Proc.\ Roy.\ Soc.\ Lond.\ A {\bf 360}, 117 (1978);
A.~D.~Linde,
Phys.\ Lett.\ B {\bf 116}, 335 (1982);
A.~A.~Starobinsky,
Phys.\ Lett.\ B {\bf 117}, 175 (1982);
A.~Vilenkin and L.~H.~Ford,
Phys.\ Rev.\ D {\bf 26}, 1231 (1982).


\bibitem{Jokinen}
A.~Jokinen,
arXiv:hep-ph/0204086.


\bibitem{AD_thermal}
R.~Allahverdi, B.~A.~Campbell and J.~R.~Ellis,
Nucl.\ Phys.\ B {\bf 579}, 355 (2000)
[arXiv:hep-ph/0001122].
A.~Anisimov and M.~Dine,
Nucl.\ Phys.\ B {\bf 619}, 729 (2001)
[arXiv:hep-ph/0008058].


\bibitem{tachyonic}
G.~N.~Felder, J.~Garcia-Bellido, P.~B.~Greene, L.~Kofman, A.~D.~Linde and I.~Tkachev,
Phys.\ Rev.\ Lett.\  {\bf 87}, 011601 (2001)
[arXiv:hep-ph/0012142].



\bibitem{coherent}
B.~Garbrecht, T.~Prokopec and M.~G.~Schmidt,
arXiv:hep-ph/0304088.






\end{thebibliography}
\end{document}